\documentclass[
    twocolumn,
    aps,
    pre,
    superscriptaddress,
    longbibliography,
]{revtex4-2}

\usepackage{graphicx,color,xcolor}
\usepackage[
    colorlinks=true,
    urlcolor=blue,
    linkcolor=blue,
    citecolor=blue,
    breaklinks,
]{hyperref}

\usepackage{amsmath}
\usepackage{amsfonts}
\usepackage{amssymb}
\usepackage{braket}
\usepackage{mathtools}
\usepackage{makecell}
\usepackage[varg]{txfonts}
\usepackage[normalem]{ulem}

\def\tr{{\rm tr }}

\newcommand{\be}{\begin{equation}}
\newcommand{\ee}{\end{equation}}

\def\Ai{\text{Ai}}

\newcommand{\beq}{\begin{equation}}
\newcommand{\eeq}{\end{equation}}

\def\ket#1{| #1 \rangle}

\begin{document}

\title{%
Spectral Edge Rigidity of Quantum Chaotic States}

\author{Joaquim Telles de Miranda}
\affiliation{Centro Brasileiro de Pesquisas F\'isicas, Rua Xavier Sigaud 150, 22290-180, Rio de Janeiro, Brazil}

\author{Tobias Micklitz}
\affiliation{Centro Brasileiro de Pesquisas F\'isicas, Rua Xavier Sigaud 150, 22290-180, Rio de Janeiro, Brazil}

\date{\today} 

\begin{abstract}

We determine the distribution of fidelity susceptibility for 
chaotic eigenstates at the spectral edge of Gaussian random-matrix ensembles. 
Previous work~\cite{Miranda2025} showed that, in the unitary class, 
the characteristic susceptibility scale of edge states grows as $D^{1/3}$, 
rather than proportionally to $D$ as in the spectral bulk,
reflecting Airy-edge level rigidity.
Extending a determinant-based framework introduced for bulk states~\cite{Sierant2019}, 
we derive the universal edge distributions for both the orthogonal and unitary ensembles. 
The two symmetry classes share the scaling variable $g/D^{1/3}$
and exhibit a symmetry-dependent cubic suppression of small susceptibilities,
while their algebraic large-$g$ tails reflect the corresponding
symmetry-dependent level repulsion.
Although eigenvector statistics retain their random-matrix form throughout the spectrum, 
edge rigidity makes low-lying chaotic states parametrically less sensitive to generic 
perturbations than bulk states. Our results establish universal, symmetry-dependent 
spectral-edge fidelity-susceptibility statistics in systems whose 
chaotic dynamics extends down to the ground state.

\end{abstract}

\maketitle

\section{Introduction}

Eigenstates of Gaussian random matrix ensembles are maximally chaotic---the 
statistics of their intensities are  energy-independent Porter-Thomas distributed \cite{PorterThomas1956} 
and entanglement entropy is given by the Page value across the entire spectrum \cite{Page1993}.
This raises a natural question: If eigenstates are equally random 
and equally entangled at all energies,
can their response to generic perturbations nevertheless depend on energy? 
Recent work answers this question affirmatively \cite{Miranda2025}. 
Comparing the distributions of the fidelity susceptibility 
(FS)---a quantum-geometric measure of state 
sensitivity to perturbations---in 
the bulk and edge of the spectrum shows that the `universality class of the 
first energy-levels' differs from that of the bulk-energy states.

Specifically, the typical FS scale for generic chaotic eigenstates 
grows linearly with the Hilbert space dimension, $g_{\rm typ}\sim D$ \cite{Flynn2026,Pandey2020,Sierant2019,Penner:2020cxk}, while
the first energy-levels exhibit a parametrically reduced scale
$g_{\rm typ}\sim D^{1/3}$ \cite{Miranda2025}.
This reduction originates from 
the rigidity of the first energy-levels 
governed by Airy-edge universality~\cite{MehtaBook,Tracy1993,Vallee2010}.
It implies that Page-entangled states near the spectral edge are
 more robust against perturbations than bulk states, despite being
equally chaotic in the conventional random-matrix sense, and identifies a
regime of `rigid chaos', where maximal entanglement coexists with enhanced
stability due to unusually sparse and rigid edge level statistics.

The universal form of the FS distribution also depends on symmetries.
The previous analysis of the spectral-edge state-rigidity focused on systems
with broken time-reversal symmetry, described by the Gaussian unitary ensemble (GUE),
where a remarkable form of semiclassical exactness allows for a controlled
field-theoretical derivation of the edge-state FS-distribution \cite{Miranda2025}.
This semiclassical structure, however, is specific to the unitary class
and does not straightforwardly extend to time-reversal invariant systems.

Motivated by the recent determinant-based approach of Sierant \textit{et al.}~\cite{Sierant2019}, 
who derived analytic fidelity susceptibility distributions for Gaussian random 
matrix ensembles in the spectral bulk—both with time-reversal symmetry 
(GOE) and without it (GUE)—we here formulate an alternative derivation 
of the FS distribution at the spectral edge.
This approach does not rely on semiclassical exactness and therefore
extends naturally to systems with time-reversal symmetry,
providing a unified framework for both orthogonal and unitary ensembles.

Our main result is the full universal FS distribution for eigenstates
in the Airy edge regime of the GOE at large Hilbert-space dimension $D\gg1$. 
We find that the edge distribution exhibits the same subextensive
scaling variable $x=g/D^{1/3}$ as in the unitary case,
but with symmetry-dependent rigidity:
exponential suppression of small $g$-values, 
with a coefficient proportional to the Dyson symmetry index 
 $\beta$, and a large-$g$ tail broadened for GOE relative to GUE,
reflecting the weaker level repulsion characteristic of the 
orthogonal class \cite{MehtaBook,haake1991quantum}.

The rest of the paper is organized as follows.
In Sec.~\ref{sec:QGT} we briefly review the
quantum geometric tensor and define fidelity susceptibility for Gaussian random
ensembles. In Sec.~\ref{sec:FS} we derive a determinant-function
representation of the FS distribution valid for both symmetry classes and evaluate it 
to leading order in $D\gg 1$ in the bulk and at the spectral edge, 
obtaining closed analytic expressions for
the GOE and GUE FS distributions. In Sec.~\ref{sec:Numerics} we test our predictions against
numerical simulations. We conclude in Sec.~\ref{sec:Discussion} with a discussion of implications
for quantum systems whose low-lying states remain chaotic down to the
spectral edge.

\section{Quantum geometric tensor}
\label{sec:QGT}

Consider a family of Hamiltonians $H(\boldsymbol{\lambda})$
 continuously depending on $k$ parameters 
$\boldsymbol{\lambda} = (\lambda_1,\dots,\lambda_k)$.
The corresponding family of eigenstates 
$\{\ket{n(\boldsymbol{\lambda})}\}$ define a metric structure
in parameter space through the line element
\begin{align}
ds^2 
= 
\sum_{\alpha\beta}
\mathrm{Re}\, g_{\alpha\beta}^{(n)}(\boldsymbol{\lambda})\,
d\lambda_\alpha d\lambda_\beta,
\end{align}
with the quantum geometric tensor (QGT)~\cite{Provost1980},
\begin{align}
\label{eq:QGT}
g_{\alpha\beta}^{(n)}
=
\langle \partial_\alpha n | \partial_\beta n \rangle
-
\langle \partial_\alpha n | n \rangle
\langle n | \partial_\beta n \rangle.
\end{align}

The real symmetric part of Eq.~\eqref{eq:QGT} describes the quantum metric,
while its imaginary antisymmetric part defines the Berry curvature~\cite{Berry1984}.
The latter vanishes in systems with time-reversal symmetry (TRS)
and is generally nonzero when TRS is broken.
Physically, $ds^2$ governs the leading quadratic decrease of the overlap 
$|\langle n(\boldsymbol{\lambda}) 
| n(\boldsymbol{\lambda}+d\boldsymbol{\lambda}) \rangle|^2$ 
under an infinitesimal parameter change.

In the following we consider the case where both the Hamiltonian
$H_0$ and a single control-parameter perturbation $H_\lambda$ ($k=1$)
are drawn independently from a Gaussian random ensemble,
\begin{equation}\label{eq:gaussian_ensembles_dist}
P(H) = \mathcal{N} \exp\!\left(-\frac{D\beta}{4}\mathrm{Tr}\,H^2\right),
\end{equation}
where for $\beta=1$ the matrices are $D$-dimensional real symmetric (GOE),
and for $\beta=2$ they are complex Hermitian (GUE), and $\mathcal{N}$ is a normalization constant.
For the present case of a single external perturbation direction
the QGT reduces to a scalar, which is identified as the fidelity susceptibility \cite{You2007}. 
Our goal here is to focus on 
large Hilbert space dimensions $D\gg1$
and determine its probability distribution
at the spectral edge.

Let us first specify what we mean by the \textit{spectral edge}. 
In the limit $D\to\infty$, the density of states (DoS) of the 
Gaussian random ensembles is described by Wigner's semicircle 
law \cite{Wigner1955,Wigner1958}, whose support, 
with the normalization of Eq.~\eqref{eq:gaussian_ensembles_dist}, 
extends over $E\in[-2,2]$ for all Dyson indices. 
The semicircle law, however, captures only the macroscopic density 
profile and does not resolve the universal fine structure appearing 
within a distance of order $D^{-2/3}$ from either endpoint. 
Near the lower edge,
\begin{equation}
E=-2+\epsilon,
\qquad
\epsilon=\mathcal{O}(D^{-2/3}),
\end{equation}
the average DoS assumes its universal Airy scaling form, 
while correlations between nearby levels are governed by the corresponding 
Airy kernel. 
This scaling region may equivalently be viewed as the critical regime of a 
symmetry-breaking transition in the retarded--advanced sector: at mean-field 
level, the DoS, proportional to the discontinuity between retarded and 
advanced Green functions, vanishes outside the spectral support and becomes 
nonzero inside it. For invariant random-matrix ensembles, the universal 
effective theory governing this critical region is the Kontsevich matrix
model~\cite{Kontsevich1992,Altland2021}.
We henceforth refer to this $D^{-2/3}$ scaling window as the \textit{spectral edge}.

\section{Fidelity susceptibility for Gaussian random ensembles}
\label{sec:FS}

Building on the general framework developed in Refs.~\cite{Sierant2019,Penner:2020cxk},
applicable to both symmetry classes, we parametrize the Hamiltonian as
$H(\lambda) = H_0 + \lambda H_\lambda$,
where $H_0$ and $H_\lambda$ are drawn independently from the same Gaussian ensemble.
For a single perturbation parameter $\lambda$, the fidelity susceptibility
of an eigenstate $\ket{n}$ of the unperturbed Hamiltonian $H_0$ 
can be expressed as 
\begin{align}
\label{eq:FS}
g_{\lambda\lambda}^{(n)} 
= 
\sum_{m \neq n} 
|\langle n | H_\lambda | m \rangle|^2\Delta_{nm}^{-2},
\end{align}
where $\Delta_{nm}\equiv E_n-E_m$, and the sum runs over all other
eigenstates $\ket{m}$ of $H_0$.

Eq.~\eqref{eq:FS} follows directly from 
Eq.~\eqref{eq:QGT} using first-order perturbation theory,
$\ket{\partial_\lambda n}
=
\sum_{m\neq n}
\ket{m}\langle m|H_\lambda|n\rangle \Delta_{nm}^{-1}$,
together with the orthogonality of the eigenstates.  
The FS thus depends on both the eigenvalues and eigenstates of the system.
As emphasized earlier, their interplay is essential in shaping
the perturbative response of quantum chaotic systems.
In particular, each term in Eq.~\eqref{eq:FS} scales as
$\Delta_{nm}^{-2}$,
making the FS highly sensitive to small level spacings. 
It is therefore directly sensitive to level repulsion,
a hallmark of quantum chaos,
whose strength depends on the underlying symmetry class.
In particular, level repulsion is more prominent when
the system lacks time-reversal symmetry~\cite{haake1991quantum}.
At the spectral edge, where level spacings follow Airy universality
and exhibit enhanced rigidity, this sensitivity produces
parametrically reduced scaling of the FS relative to bulk states.

Introducing then the probability distribution for the FS at energy $E$ 
\begin{align}
P_E(g) 
&\propto
\Bigg\langle \sum_n \delta(E - E_n)\delta(g-g_{\lambda\lambda}^{(n)}) \Bigg\rangle_{H_0, H_\lambda},
\end{align}
we insert Eq.~\eqref{eq:FS} and take the Fourier transform
to obtain the characteristic function.
Averaging the latter over the perturbation $H_\lambda$ 
with respect to the distribution in Eq.~\eqref{eq:gaussian_ensembles_dist}
is straightforward. This leads to the expression,
\begin{align}
\label{eq:intermediate_FT_representation}
P_E(\omega) 
&= \left\langle \sum_{n} \delta(E - E_n) \prod_{m\neq n} 
    \Delta_{mn}^{\beta}
    \left( \Delta_{mn}^2 - \frac{2 i\omega
    }{ 
        \beta D}\right)^{-\frac{\beta}{2}}
        \right\rangle_{H_0},
\end{align}
where The Dyson-index dependence can be traced back to the
fact that the GOE consists of real matrices, while the GUE's elements
are complex matrices. Further progress is made 
using the joint eigenvalue distribution for $H_0$ 
~\cite{VonOppen1994}, and a simple calculation leads to the compact formula,
\begin{align}
\label{eq:characteristic_form}
P_E(\omega) 
&= 
\Bigg
\langle
    \frac{\det( H_E)^{2\beta}}{
        \prod_{j}^2 \det(H_E + (-1)^j a)^{\beta/2}}
  \Bigg
    \rangle_{H}.
\end{align}
Here $a= e^{i\pi{\rm sgn}(\omega)/4}\sqrt{2 |\omega|/\beta D}$, 
and $H_E\equiv H-E$ with $H$ a $(D-1)$-dimensional reduced matrix from the Gaussian ensemble.

Eq.~\eqref{eq:characteristic_form} reduces the calculation to an ensemble
average of a ratio of characteristic polynomials, a class of objects that has
been studied by a variety of methods
\cite{VonOppen1994,VonOppen1995,Fyodorov1995,Fyodorov2011,Fyodorov2012,Poli2009}.
To evaluate this average in a form applicable to both the orthogonal and
unitary symmetry classes, and amenable to the spectral-edge scaling limit,
we adapt the determinant-based approach of Sierant \textit{et al.}~\cite{Sierant2019}.
We first represent the determinant in the denominator as a Gaussian integral
over a $(D-1)$-dimensional real vector for $\beta=1$, or complex vector
field for $\beta=2$, using 
\begin{align}
    \label{eq:functional_rep}
\left(\det\left(H_E^2 - a^2\right)\right)^{-\frac{\beta}{2}} 
&= 
\int d(\mathbf{z}^\dagger\mathbf{z}) 
\, e^{-\mathbf{z}^\dagger 
\left(
    H_E^2 - a^2 
\right)\mathbf{z}},
\end{align}
where $d(\mathbf{z}^\dagger\mathbf{z}) $ denotes the normalized Gaussian measure over
real ($\beta=1$) or complex ($\beta=2$) $(D-1)$-component vectors,
such that $\int d(\mathbf{z}^\dagger\mathbf{z}) \, e^{-\mathbf{z}^\dagger \mathbf{z}}=1$.
The integrand inherits the symmetry 
of the underlying Gaussian matrix ensemble, i.e. is 
invariant under transformations 
under the orthogonal, respectively, unitary group. 
 This motivates one to express $\mathbf{z} = r{\bf e}_1$, 
with $\mathbf{e}_1$ the unit vector along the first direction,
leaving thus only 
the nontrivial one-dimensional 
 $r$ integral. 
The inverse Fourier 
transform, $P_E(g) \propto \int d\omega\, e^{- i \omega g} P_E(\omega)$, 
generates a $\delta$-function for $g$, 
resulting in
\begin{align}
    \label{eq:intermediate_distribution}
P_E(g) 
&\propto 
\int_0^\infty dr\, r^{\beta(D-1)-1} 
\delta\left(D \beta g - 2 r^2\right)
\langle F(H_E) \rangle_H,
\end{align}
where we introduced
\begin{align}
\label{eq:F_H}
F(H)
&=
{\rm det}(H)^{2\beta}\, e^{-r^2\sum_{j = 1}^D |H_{1j}|^2}. 
\end{align}

To perform the $H$-average of Eq.~\eqref{eq:F_H}, 
we parametrize 
$H= 
\left(
    \begin{smallmatrix}
h  && {\bf x} \\
{\bf x}^\dagger && V  \\
\end{smallmatrix}\right)$, 
with $h=H_{11}$ a scalar, 
 $V$ the $(D-2)$-dimensional submatrix obtained by removing
the first row and column of $H$,
and ${\bf x}$ the $(D-2)$-component vector
formed by the first row, excluding $H_{11}$.
 Invoking then the identity
$\det
\left(
\begin{smallmatrix}
A & B \\
C & D \\
\end{smallmatrix}
\right)
=
\det(A - B D^{-1}C) \det(D)$, 
the average of Eq.\eqref{eq:F_H} can be recast as
\begin{align}
    \label{eq:intermediate_determinant_avg}
&\langle 
F(H_E)
\rangle_{H} 
\propto 
e^{- cE^2r^2} \left(2a\right)^{-1/2}\left(2b\right)^{-(D-2)} 
\times 
\nonumber \\ 
&\times  
\left\langle\left(
    \frac{1}{\sqrt{2a}}h 
    - cE 
    - \frac{1}{2b}\mathbf{x}^\dagger \left(V_E^{-1}\right)\mathbf{x}
    \right)^{2\beta}\det\left( V_E
\right)^{2\beta}    \right\rangle,
\end{align}
where $V_E \equiv V - E$, 
\begin{align}
a \equiv \frac{\beta D}{4} + r^2, \quad 
b \equiv \frac{\beta D}{2} + r^2, \quad 
c \equiv \frac{\beta D}{\beta D + 4 r^2}.
\end{align}
The average is taken over the Gaussian variables $h$, $\mathbf{x}$, and $V$, with
$\sigma_h^2=\sigma_{\mathbf{x}}^2=1$ and
$\sigma_V^2=\sqrt{2/(D\beta)}$, as in
Eq.~\eqref{eq:gaussian_ensembles_dist}; see
Appendix~\ref{app:derivation_of_inermediate_result} for details~\footnote{
We retain the $D$-dependent normalization of the original ensemble even
though $V$ has dimension $(D-2)\times(D-2)$, i.e.,
$P(V)\propto\exp[-(D\beta/4)\tr V^2]$.}. 
Up to this point, the derivation applies to both symmetry classes. We now
evaluate the remaining Gaussian integrals separately for the orthogonal and
unitary ensembles.

\begin{figure*}[t!]
\centering
\includegraphics[width=.48\linewidth]{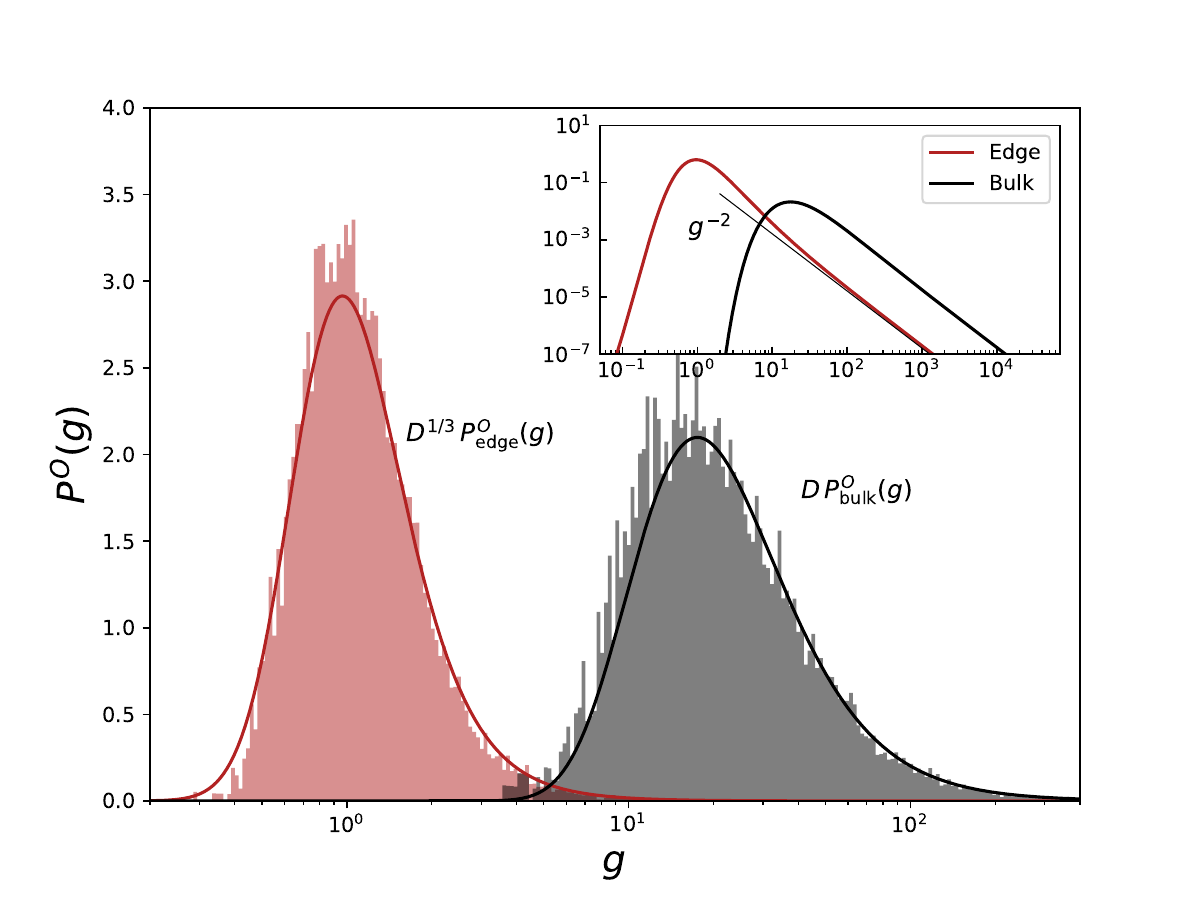}
\includegraphics[width=.48\linewidth]{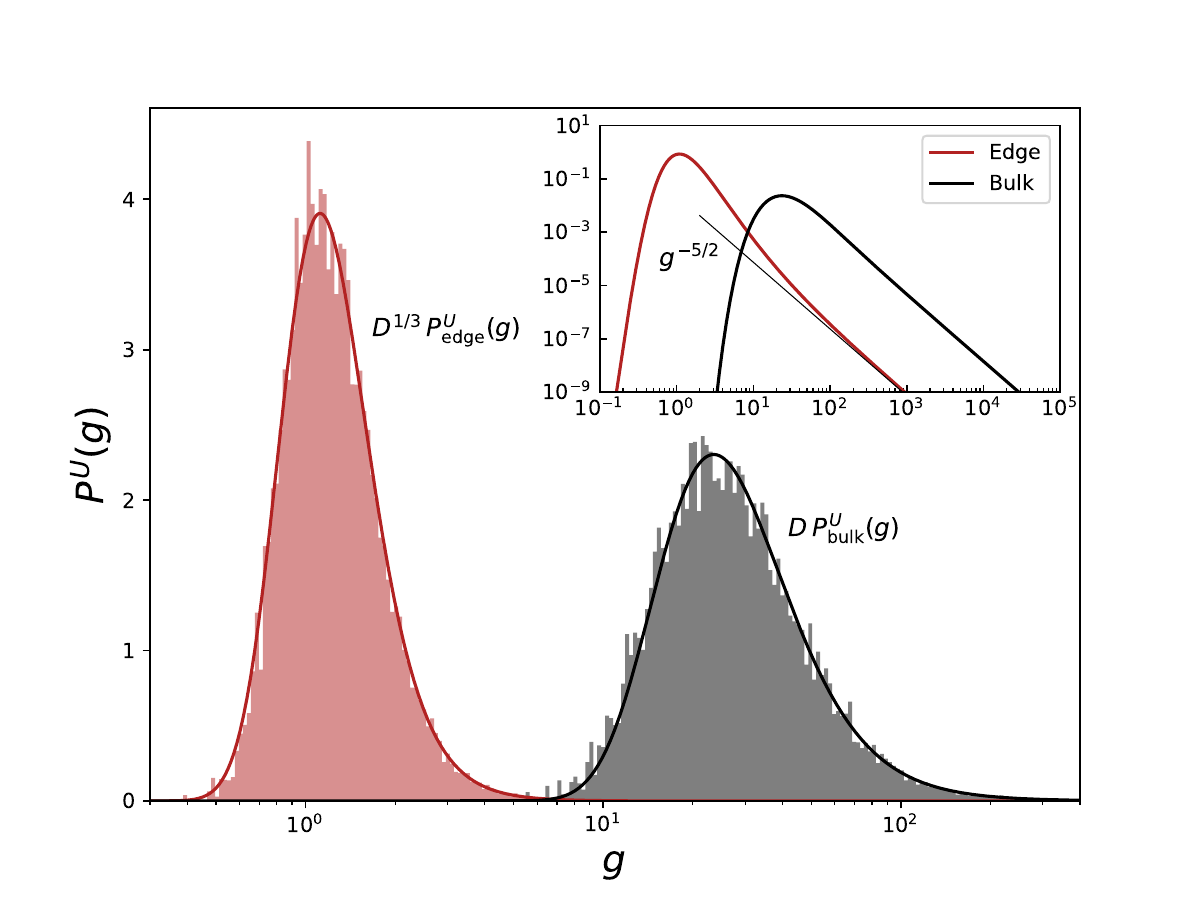}
\caption{
Left: GOE fidelity susceptibility distributions for near-edge 
(red) and bulk (black) states compared with histograms generated from an 
ensemble of $10{,}000$ matrices of dimension $D=100$. 
(Edge and bulk distributions are scaled by $D^{1/3}$ 
and $D$, respectively, for visibility.) 
Inset: GOE distributions for near-edge and bulk states, 
with asymptotic scaling $g^{-2}$ shown for reference. 
Right: Corresponding plot for the GUE, with the inset now 
displaying $g^{-5/2}$ scaling. As noted in the main text, 
edge distributions were shifted as 
$P^{O,U}_{\rm edge}(g - g^*)$, with 
$g^* \approx 0.50, 0.47$ for GOE and GUE, respectively.
}
\label{fig:numerics_ch2}
\end{figure*}

\subsection{GOE}

For GOE, $\beta = 1$, we can work with real symmetric matrices, 
simplifying the subsequent calculations. The
integral over the scalar $h$-component is readily
done,
$\left\langle
    \left( h/\sqrt{2a} - X \right)^2 \right\rangle_h
    \propto 
    1/(2 a) + X^2$,
and integration over remaining components is 
simplified by orthogonal invariance of $V$. 
Diagonalizing the latter, we simplify
\begin{align}
    \label{eq:GOE_moments} 
&\langle
(\mathbf{x}^T V_E^{-1} \mathbf{x})^n\det 
V_E^2 
\rangle_{\mathbf{x},V}
 \nonumber \\ 
&= 
\begin{cases}
\left\langle \det  V_E^2 \right\rangle_V &\,\, n = 0, \\
\left\langle \tr V_E^{-1} \det V_E^2 \right\rangle_V &\,\, n = 1, \\
\left\langle \left[(\tr V_E^{-1})^2 + 2\tr V_E^{-2}\right] 
\det V_E^2  \right\rangle_V &\,\, n = 2.
\end{cases}
\end{align}
We then perform the remaining $V$-integral 
with help of the identity
$\partial_j \det(V - j) = - \det(V - j)\tr(V - j)^{-1}$, 
introducing the generating function
\begin{align}
    \label{eq:GOE_generating_function}
&Z^O 
= 
\left\langle\det(V - j_1)\det(V - j_2)\right\rangle_V.
\end{align}
Expressing the averages further as
\begin{align}
    \label{eq:GOE_moments_def}
I_1^O
&\equiv
-\frac{\partial_{j_1}Z^O}{Z^O}\Bigg|_{j_i = E}
=
\frac{\left\langle \tr V_E^{-1} \det V_E^2 \right\rangle_V}{\left\langle \det V_E^2 \right\rangle_V},
\nonumber \\
I_2^O
&\equiv
\frac{(3\partial_{j_1}\partial_{j_2} - 2\partial_{j_1}^2)
Z^O}{Z^O}\Bigg|_{j_i = E}= 
\frac{\left\langle \left[(\tr V_E^{-1})^2 + 2\tr V_E^{-2}\right] \det V_E^2 \right\rangle_V}{\left\langle\det V_E^2 \right\rangle_V},
\end{align}
 leads to the FS distribution 
\begin{widetext}
\begin{align}
P^O(g) &\propto \int_0^\infty dr\, 
\delta\left(g - \frac{2 r^2}{D}\right) 
r^{D - 2} e^{-c E^2 r^2} (2a)^{-1/2} 
    (2b)^{-(D - 2)} 
\left(
    \frac{1}{2a} + c^2 E^2 + \frac{c E}{b} I_1^O + \frac{1}{4b^2}I_2^O
    \right) 
    \nonumber \\
&\propto 
e^{-D f^O_E(g)} \frac{1 + g}{g^{3/2}\sqrt{1 + 2 g}} 
\left(
    \frac{2}{D(1 + 2g)} 
    + 
    \frac{E^2}{(1 + 2g)^2} 
    + 
    \frac{2 E}{D(1 + 2g)(1 + g)} I_1^O 
    + 
    \frac{1}{D^2(1 + g)^2}I_2^O
    \right),
\end{align}
\end{widetext}
with $f_E^O(g) = E^2 g / (2 + 4 g) + (1/2)\ln\left(1 + 1 / g\right)$. 

    The evaluation of the moments $I_j^O$ depends on the spectral regime.
Although exact finite-$D$ expressions for such characteristic-polynomial
moments are available~\cite{Delannay2000}, here we retain only their leading
large-$D$ behavior, using the results of Ref.~\cite{Brezin2001}.
For fixed energies in the spectral bulk,
$|E|<2$ with $2-|E|=\mathcal{O}(1)$, the saddle-point evaluation of
$Z^O$ described in Appendix~\ref{app:goe_generating_function} gives 
\begin{align}
\label{eq:general_result_bulk_GOE}
P^O_{\rm bulk}(g) 
&\propto
p\left(g/(D\rho_E^2)\right)
e^{-\frac{D\rho_E^2}{2g}}, \nonumber \\
p(x) &= x^{-2}  +  x^{-3}, 
\end{align}
with 
$\rho_E = \sqrt{1-E^2/4}$ 
the density of states, here normalized to $\rho_0=1$.  
This  
generalizes previous results of Ref.~\cite{Sierant2019} 
for $E = 0$ to arbitrary energies in the spectral bulk.

Our main focus here is on eigenstates at the edge,
and the energy is now measured relative to the lower spectral edge as
$\epsilon = E + 2$.
The edge analysis below concerns eigenstates with
$\epsilon=\mathcal{O}(D^{-2/3})$, i.e., throughout the Airy scaling window,
where the generating function is governed by Airy-kernel universality.

As previously stated,
in this regime the generating function
(\ref{eq:GOE_generating_function})
is governed by Airy-kernel universality~\cite{Vallee2010},
and the determinant moments $I_j^O$ can be expressed
in terms of Airy functions and their derivatives.
Evaluating these to leading order in $D\gg1$, as detailed in Appendix~\ref{app:goe_generating_function},  
we obtain the edge distribution
\begin{align}
\label{eq:result_edge_GOE}
P^O_{\rm edge}(g) 
&\propto
p\left(g/D^{1/3}\right)
e^{-\frac{D}{24 g^3}}, \nonumber \\
p(x) &= c_2 x^{-2} + c_3 x^{-3}
+ c_4 x^{-4} + c_5 x^{-5},
\end{align}
with numerical coefficients
$c_2\approx 1$, $c_3\approx 2.06$, $c_4\approx 1.09$, and $c_5\approx 0.25$.
Exact expressions are rather cumbersome, and can be readily derived 
from the results explicited in Appendix~\ref{app:goe_generating_function}.

Although the full distribution is modified at the spectral edge, its large-$x$
asymptotics remain $p(x)\sim x^{-2}$,
for $x\gg1$, 
with the same exponent as in the bulk. This common exponent reflects the
dominance of rare, anomalously small spacings between neighboring levels,
whose level-repulsion behavior is unchanged at the edge. 
However, the overall scale of fluctuations is parametrically reduced.
The scaling variable $x=g/D^{1/3}$ implies that typical susceptibilities
scale as $g_{\rm typ}\sim D^{1/3}$ rather than $D$.
Moreover, the exponential factor
$\exp\left(-D/(24 g^3)\right)$ produces a stronger suppression
of small-$g$ fluctuations than in the bulk,
signaling enhanced rigidity of Airy-correlated edge levels.
As a result, the full distribution is significantly compressed
relative to bulk states, see Fig.~\ref{fig:numerics_ch2}.

\subsection{GUE}

For GUE, $\beta=2$, we proceed analogously, though with increased complexity. 
Relevant moments are now generated from the function
\begin{align}\label{eq:GUE_generating_function}
Z^U = \left\langle \prod_{j=k}^4 \det(V - j_k) 
\right\rangle_V,
\end{align}
involving four sources $\{j_k\}_{k=1,...,4}$. 
For the unitary ensemble, we use the leading large-$D$ expression for
$Z^U$ derived in Ref.~\cite{Brezin2000}, exact finite-$D$ results are
available in Refs.~\cite{Strahov2003,Fyodorov2003}. Its asymptotic
evaluation depends on the spectral regime. In the bulk, it reproduces
the known result of Ref.~\cite{Andreev1995}, whereas in the
$D^{-2/3}$ edge-scaling regime it assumes a universal form governed by
the Airy kernel. The derivation is given in
Appendix~\ref{app:gue_generating_function}, here we state only the
result.

Starting out from the distribution for general energies $E$,
\begin{widetext}
\begin{align}
P^U(g) 
\propto& 
e^{-D f^U_E(g)} \frac{(1 + g)^2}{g^2\sqrt{1 + 2g}} 
\Bigg(
    \frac{3}{D^2(1 + 2g)^2} 
    + 
    \frac{6E^2}{D(1 + 2g)^3} 
    + 
    \frac{E^4}{(1 + 2g)^4} 
    + 
    \left[
        \frac{12 E}{D^2(1 + g)(1 + 2g)^2} + \frac{4 E^3}{D(1 + g)(1 + 2g)^3}
        \right]I^U_1 
        \nonumber \\ 
&\qquad\qquad\qquad
+ 
\left[
    \frac{6}{D^3(1 + g)^2(1 + 2g)} 
    + 
    \frac{6 E^2}{D^2(1 + g)^2(1 + 2g)^2}
    \right]I^U_2 
    + 
    \frac{4E}{D^3(1 + g)^3(1 + 2g)} I^U_3 
    + 
    \frac{1}{D^4(1 + g)^4} I^U_4 
    \Bigg),
\end{align}
\end{widetext}
where $f^U_E(g) = g E^2 / (1 + 2 g) + \ln(1 + 1/g)$, 
we first focus on energies in the bulk. 
Restricting to  $|E| < 2$ 
away from the spectral edge, 
we find
\begin{align}\label{eq:general_result_bulk_GUE}
P^U_{\rm bulk}(g) 
&\propto
    p\left(g/(D\rho_{E}^2)\right)
    e^{-\frac{D\rho_{E}^2}{g}}, \nonumber \\ 
p(x) &= \frac{3}{4}x^{-5/2} + x^{-7/2} + x^{-9/2},
\end{align}
in agreement with previous work~\cite{Sierant2019,Penner:2020cxk,Miranda2025}.

For eigenstates at the spectral edge, $\epsilon = E + 2$ with
$\epsilon = \mathcal{O}(D^{-2/3})$,
 we obtain
\begin{align}
    \label{eq:result_edge_GUE}
P^U_{\rm edge}(g) 
&\propto  
p\left(g/D^{1/3}\right) e^{-\frac{D}{12 g^3}}, \nonumber \\
p(x)&= x^{-5/2}+k_7 x^{-7/2}+ \dots + k_{17}x^{-17/2},
\end{align}
with coefficients 
$k_7\approx 7.12$, $k_9\approx 11.61$, $k_{11}\approx 8.72$, 
$k_{13}\approx 3.56$, $k_{15}\approx 0.79$, and $k_{17}\approx 0.08$, 
and in agreement with previous work~\cite{Miranda2025}.
Exact expressions can be derived following Appendix~\ref{app:gue_generating_function}.

A comparison of the spectral-edge distributions for the orthogonal,
Eq.~\eqref{eq:result_edge_GOE}, and unitary,
Eq.~\eqref{eq:result_edge_GUE}, ensembles reveals a common scaling
structure with symmetry-class-dependent algebraic prefactors. Both
distributions involve the subextensive scaling variable
$x=g/D^{1/3}$ and exhibit the cubic exponential suppression
$\exp[-\mathcal{O}(D/g^3)]$, but differ in their algebraic dependence
on $x$. 
For the GOE the leading tail behaves as $p(x)\sim x^{-2}$,
whereas for the GUE it decays more rapidly as
$p(x)\sim x^{-5/2}$.
This steeper decay reflects the stronger level repulsion
in the unitary class $\beta=2$,
which suppresses near-degeneracies and therefore reduces
the probability of large fidelity-susceptibility fluctuations.
In this sense, edge states in the GUE are statistically
more rigid than those in the GOE.
Finally, the coefficient of the cubic exponential suppression scales
proportionally to the Dyson index $\beta$,
demonstrating explicitly how symmetry controls
spectral-edge rigidity.

\section{Numerical validation}
\label{sec:Numerics}

To validate our analytical results, we compare them to numerical
simulations of Gaussian random matrices using exact diagonalization.
For each realization, we diagonalize $H_0$ and compute the fidelity
susceptibility from Eq.~\eqref{eq:FS}.
The distributions are obtained by averaging over $10^4$
independent matrices of dimension $D=100$.

When focusing on edge states, it is essential to select eigenvalues
within the Airy scaling window.
In the large-$D$ limit the spectral edge is located at
$E_{\rm edge}=-2$ (in our normalization),
and the Airy regime corresponds to energies satisfying
$\epsilon=E-E_{\rm edge}=\mathcal{O}(D^{-2/3})$.
In practice, we therefore pool the lowest eigenstates whose energies lie within
a narrow window $|E-E_{\rm edge}|\lesssim c\,D^{-2/3}$, with a numerical
prefactor $c$ chosen such that the statistics remain stable while staying
inside the Airy regime.

Fig.~\ref{fig:numerics_ch2} shows excellent agreement between
analytical predictions and numerical histograms,
both in the bulk and at the spectral edge.
For finite $D$, we observe a small systematic shift
$g \rightarrow g + g_0$ with $g_0=\mathcal{O}(1)$
in the edge distributions.
This shift originates from subleading $1/D$ corrections
in the determinant moments that are neglected in our
large-$D$ analysis, see Appendix~\ref{app:goe_generating_function} for further details.
Since the typical scale at the edge is
$g_{\rm typ}\sim D^{1/3}$,
the relative correction $g_0/g_{\rm typ}\sim D^{-1/3}$
vanishes in the thermodynamic limit.

\section{Conclusion}
\label{sec:Discussion}

We have derived the leading large-$D$ probability distributions of the
fidelity susceptibility for the Gaussian orthogonal and unitary ensembles,
both in the spectral bulk and in the universal Airy edge regime. Building
on the random-matrix framework of Ref.~\cite{Sierant2019}, our treatment
places the two symmetry classes within a common approach and extends
previous bulk results~\cite{Sierant2019,Penner:2020cxk} to states near the
spectral edge. Although technically more involved, the extension to the
Gaussian symplectic ensemble should require no fundamentally new ingredients
and may be pursued using the same general strategy.

A central result is the subextensive scaling of the typical fidelity
susceptibility at the spectral edge,
$g_{\rm typ}\sim D^{1/3}$, compared with
$g_{\rm typ}\sim D$ in the bulk.
This reduction does not originate from any change in eigenvector statistics,
which retain their symmetry-class-specific random-matrix form. 
Instead, it reflects the change in spectral scaling: the characteristic
level spacing near the edge is of order $D^{-2/3}$, rather than $D^{-1}$
in the bulk, and the level statistics are 
governed by Airy universality.

The FS edge distributions exhibit a symmetry-dependent
cubic exponential suppression,
$\exp\left(-\beta D/(24 g^3)\right)$,
whose coefficient scales proportionally to the Dyson index $\beta$.
The algebraic tails $\sim x^{-(3 + \beta)/2}$
further reflect symmetry-dependent 
level repulsion, leading to a stronger suppression of large 
fluctuations in the unitary class.
These results establish universal, symmetry-dependent spectral-edge
fidelity-susceptibility statistics for dense chaotic systems~\cite{Altland:2024ubs}. 
They show
that, when chaos persists down to the ground state, low-lying states are
parametrically less susceptible to generic perturbations than their bulk
counterparts.
Prominent examples of such systems include low-dimensional 
gravitational models~\cite{Mertens2023,Altland:2020ccq} and 
SYK-type theories~\cite{Cotler:2016fpe}, where spectral-edge universality may 
control the stability of low-lying states.

{\bf Acknowledgements:} T.~M. acknowledges financial support by 
Brazilian agencies CNPq and FAPERJ, and J.~T.~M. 
financial support by Brazilian agency CAPES.

\bibliographystyle{apsrev4-1}
\bibliography{bibliography}

\clearpage
\appendix

\begin{widetext}

\section{Derivation of Eq.~\eqref{eq:intermediate_determinant_avg}}\label{app:derivation_of_inermediate_result}

Starting out from Eqs.~\eqref{eq:intermediate_distribution} and \eqref{eq:F_H}, 
we decompose the $(D-1)\times(D-1)$ matrix $H$ by
separating its first row and column
\begin{align}
H=
\begin{pmatrix}
h & \mathbf{x}^{\dagger}\\
\mathbf{x} & V
\end{pmatrix},
\end{align}
where $h=H_{11}$ is a scalar, $\mathbf{x}$ is a
$(D-2)$-component vector, and $V$ is a
$(D-2)\times(D-2)$ matrix. Applying the Schur-complement identity
\begin{align}
\det
\begin{pmatrix}
A & B \\
C & D \\
\end{pmatrix}
=
\det(A - B D^{-1}C) \det(D),
\end{align}
the average in Eq.\eqref{eq:intermediate_distribution} 
takes the form
\begin{align}
    \langle &F(H_E)\rangle_H = \left\langle \left(h - E - {\mathbf x}^\dagger\left(V - E\right)^{-1} {\mathbf x}\right)^{2\beta} {\rm det}(V - E)^{2\beta}\, e^{-r^2 (h - E)^2}e^{-r^2 {\mathbf x}^\dagger {\mathbf x}} \right\rangle_H \nonumber \\
    &\propto \int dh\, e^{-\frac{D \beta}{4} h^2} \int d({\mathbf x}^\dagger{\mathbf x})\, e^{-\frac{D \beta}{2} {\mathbf x}^\dagger{\mathbf x}} \int dV\, e^{-\frac{D \beta}{4} {\rm Tr} V^2}  \left[\left(h - E - {\mathbf x}^\dagger\left(V - E\right)^{-1}{\mathbf x}\right)^{2\beta} {\rm det}(V - E)^{2\beta}\, e^{-r^2 (h - E)^2}e^{-r^2 {\mathbf x}^\dagger{\mathbf x}}\right] \nonumber \\
    &\propto \int dh\, \int d({\mathbf x}^\dagger{\mathbf x})\, \int dV\, e^{-a\left(h - \frac{r^2 E}{a}\right)^2 - c r^2 E^2}e^{-b {\mathbf x}^\dagger{\mathbf x}}e^{-\frac{D \beta}{4} {\rm Tr} V^2}\left[\left(h - E - {\mathbf x}^\dagger\left(V - E\right)^{-1}{\mathbf x}\right)^{2\beta} {\rm det}(V - E)^{2\beta}\, \right], \\
\end{align}
where the second equality follows by inserting the Gaussian measure of
Eq.~\eqref{eq:gaussian_ensembles_dist}. In the final step, we collect the
quadratic terms and introduce
\begin{align}
a \equiv \frac{\beta D}{4} + r^2, \quad 
b \equiv \frac{\beta D}{2} + r^2, \quad 
c \equiv \frac{\beta D}{\beta D + 4 r^2}.
\end{align}
Finally, we shift and rescale the integration variables according to
$h \to h/\sqrt{2a} + r^2 E/a$ and 
${\mathbf x} \to {\mathbf x}/\sqrt{2b}$. 
Including the corresponding Jacobians, we obtain
\begin{align}
\langle F(H_E)\rangle_H
&\propto
\frac{e^{-cr^2E^2}}
{\sqrt{2a}\,(2b)^{\beta(D-2)/2}}
\int dh\, e^{-h^2/2}
\int d({\mathbf x}^\dagger{\mathbf x})\,
e^{-\mathbf{x}^\dagger\mathbf{x}/2}
\int dV\,
e^{-\frac{D\beta}{4}\operatorname{Tr}V^2}
\left[
\left(
\frac{h}{\sqrt{2a}}
-cE
-\frac{1}{2b}
\mathbf{x}^\dagger(V-E)^{-1}\mathbf{x}
\right)^{2\beta}
\det(V-E)^{2\beta}
\right],
\end{align}
where $d({\mathbf x}^\dagger{\mathbf x})$ denotes integration over the
$\beta(D-2)$ independent (complex) components of $\mathbf{x}$.
Recognizing the remaining integrals as Gaussian ensemble averages
reproduces Eq.~\eqref{eq:intermediate_determinant_avg} and completes
the derivation.


\section{GOE generating function}\label{app:goe_generating_function}

The generating function for GOE introduced in  
Eq.~\eqref{eq:GOE_generating_function} 
in the main text
\begin{align*}
Z^O 
&= 
\left\langle\det(V - j_1)\det(V - j_2)\right\rangle_V,
\end{align*}
 can be evaluated by the saddle point method at large $D$. 
Using a generalization of the Harish-Chandra-Itzykson-Zuber
method to the orthogonal ensemble, Brézin and Hikami derived the
integral representation~\cite{Brezin2001}
\begin{align}
\label{app_eq:intermediate_GOE}
Z^O
&=
e^{-\frac{D}{2}(j_1^2+j_2^2)}
\int_{-\infty}^{\infty}dt_1dt_2\,
(t_1t_2)^D
e^{-D(t_1^2+t_2^2)
+\sqrt{2}iD(t_1j_1+t_2j_2)}
\left[
\frac{2}{D}
\left(
\frac{t_1-t_2}{j_1-j_2}
\right)^2
+
\frac{2\sqrt{2}i}{D^2}
\frac{t_1-t_2}{(j_1-j_2)^3}
\right].
\end{align}
The corresponding saddle points are
\begin{align}
\bar t_{a,\pm}
=
\frac{1}{\sqrt{2}}
\left(
\frac{ij_a}{2}
\pm
\sqrt{1-\frac{j_a^2}{4}}
\right),
\qquad a=1,2.
\end{align}
For real $j_a$ in the spectral bulk, their real parts contain the
semicircle factor,
$\operatorname{Re}\bar t_{a,\pm}
=
\pm\frac{\pi}{\sqrt{2}}\rho_{\rm sc}(j_a)$,
with 
$\rho_{\rm sc}(j)
=
\frac{1}{2\pi}\sqrt{4-j^2}$.

For fixed energies in the spectral bulk, the integral in
Eq.~\eqref{app_eq:intermediate_GOE} can be evaluated by the
saddle-point method. Of the four possible saddle-point combinations,
only two contribute at leading order in the large-$D$ limit; the
remaining two are suppressed by a relative factor of order $D^{-1}$.
Retaining the leading contributions yields the bulk result of
Ref.~\cite{Brezin2001},
\begin{align}
\label{app_eq:bulk_generating_function}
Z^O_{\rm bulk}
&\propto
e^{\frac{D}{4}(j_1^2+j_2^2)}
\left(
\frac{\cos x}{x^2}
-
\frac{\sin x}{x^3}
\right),
\end{align}
where
$x = D(j_1 - j_2)\sin\left[(\arccos\left(j_1/2\right) 
+ \arccos\left(j_2/2\right))/2\right]$.

The bulk saddle-point expansion ceases to be uniform as the spectral edge is
approached. At $j_a=-2$, the real parts of the two saddle points vanish and
the saddles coalesce, so that the Gaussian expansion used in the bulk is no
longer sufficient. We therefore return to
Eq.~\eqref{app_eq:intermediate_GOE} and rewrite its algebraic prefactor in
terms of derivatives with respect to $j_1$ and $j_2$
\begin{align}
Z^O 
&= 
e^{\frac{D}{2}(j_1^2 + j_2^2)} 
\left[
    \frac{1}{D}\frac{1}{(\sqrt{2} i D)^2}
    \left(\frac{\partial_{j_1} - \partial_{j_2}}{j_1 - j_2}
    \right)^2 
    + 
    \frac{i}{D}\frac{1}{\sqrt{2} i D}\frac{\partial_{j_1} - \partial_{j_2}}{(j_1 - j_2)^3}
    \right] 
    \nonumber \\ 
&\qquad\qquad\qquad\qquad\qquad
\times
\int_{-\infty}^\infty dt_1dt_2\, 
\exp\left[D\sum_{a = 1, 2}
\left(
    -t_a^2 + \sqrt{2}it_aj_a + \ln t_a
    \right)
    \right].
\end{align}
To extract the edge scaling limit, we measure the source variables relative
to the lower spectral edge by setting $j_a\to-2+j_a$ and expand the exponent
about the coalescing saddle point. Keeping the leading nonvanishing terms
gives
\begin{align}\label{app_eq:edge_generating_function}
Z_{\rm edge}^O 
&= 
e^{\frac{D}{2}(j_1^2 + j_2^2) - 2D(j_1 + j_2)} 
\left[
    -\frac{1}{2D^3}\left(\frac{\partial_{j_1} 
    - \partial_{j_2}}{j_1 - j_2}\right)^2 
    + \frac{1}{\sqrt{2}D^2}\frac{\partial_{j_1} 
    - \partial_{j_2}}{(j_1 - j_2)^3}
    \right]
    \times \nonumber \\ 
&\qquad\qquad\qquad\qquad\qquad
\times 
e^{D(j_1 + j_2)}
\int_{-\infty}^\infty dx_1dx_2\, 
\exp
\left[
    iD\sum_{a = 1, 2}
    \left(
        -\frac{2\sqrt{2}}{3}x_a^3 + \sqrt{2}j_ax_a
        \right)
        \right] 
        \nonumber \\
&= 
e^{\frac{D}{2}(j_1^2 + j_2^2) 
- 
2D(j_1 + j_2)} 
\left[
    -\frac{1}{2D^3}
    \left(
        \frac{\partial_{j_1} 
        - 
        \partial_{j_2}}{j_1 - j_2}\right)^2 
        + 
        \frac{1}{\sqrt{2}D^2}\frac{\partial_{j_1} 
        - 
        \partial_{j_2}}{(j_1 - j_2)^3}
        \right]
        \times 
        \nonumber \\ 
&\qquad\qquad\qquad\qquad\qquad
\times 
e^{D(j_1 + j_2)}
\Ai\left(-D^{2/3}j_1\right)
\Ai\left(-D^{2/3}j_2\right) 
\nonumber \\ 
&= 
e^{\frac{D}{2}(j_1^2 + j_2^2) 
-D(j_1 
+ j_2)} 
\frac{\partial_{j_1} 
- \partial_{j_2}}{j_1 - j_2
}
\times
\nonumber \\ 
&\qquad
\times
\left[
    \frac{\Ai\left(-D^{2/3}j_1\right)
    {\rm Ai}'\left(-D^{2/3}j_2\right)
    -
    {\rm Ai}'\left(-D^{2/3}j_1\right)
    \Ai\left(-D^{2/3}j_2
    \right)}{j_1 - j_2}\right].
\end{align}
Equation~\eqref{app_eq:edge_generating_function} retains the leading terms
in the large-$D$ edge-scaling limit. The cubic expansion of the exponent
implies $x_a=\mathcal{O}(D^{-1/3})$, and 
$j_a=\mathcal{O}(D^{-2/3})$,
so that all higher-order terms in the saddle-point expansion are
parametrically suppressed. The resulting integrals therefore reduce to
Airy functions, yielding the universal edge form displayed above. 

    Substituting the bulk and edge expressions,
Eqs.~\eqref{app_eq:bulk_generating_function} and
\eqref{app_eq:edge_generating_function}, into the moment formula
Eq.~\eqref{eq:GOE_moments_def} yields the corresponding fidelity-susceptibility
distributions. In the edge-scaling regime, this procedure gives
Eq.~\eqref{eq:result_edge_GOE}, with algebraic prefactor
 $p(x) \approx x^{-2}  +  2.06\,x^{-3} +  1.09\, x^{-4}  +  0.25\, x^{-5}$.

\section{GUE generating function}\label{app:gue_generating_function}

The analysis of the GUE generating function closely parallels the GOE
calculation, with two principal differences: the generating function now
contains four determinants, while its large-$D$ integral representation is
simpler than in the orthogonal case. We therefore omit steps that repeat the
preceding derivation and state the results needed below.

Following Brézin and Hikami~\cite{Brezin2000}, the unitary generating
function defined in Eq.~\eqref{eq:GUE_generating_function} admits the
integral representation
\begin{align}
\label{app_eq:intermediate_GUE}
Z^U
&\propto
e^{-\frac{D}{2}\sum_{i = 1}^4 j_i^2}
\int_{-\infty}^\infty \prod_{i = 1}^4
\left(dt_i\,t_i^D\right)
e^{-\frac{D}{2}\sum_{i = 1}^4(t_i^2 + 2it_ij_i)}
\prod_{m < n}\frac{t_m - t_n}{j_m - j_n}.
\end{align}

For fixed energies in the spectral bulk, we evaluate this integral by
summing the leading saddle-point configurations satisfying
${\rm Re}(t_1+t_2+t_3+t_4)=0$. This yields
\begin{align}
\label{app_eq:generating_function_bulk_GUE}
Z^U_{\rm bulk}
&\propto
e^{-D \sum_{i = 1}^4j_i^2}
\sum_{\sigma \in S_4} e^{-D\sum_{i=1}^4
\left(
e^{- (-1)^i\theta_{\sigma(i)}} - i\theta_{\sigma(i)}
\right)}
\prod_{m<n}\frac{e^{-i(-1)^m\theta_{\sigma(m)}}
-
e^{-i(-1)^n\theta_{\sigma(n)}}}{
\cos\left(\theta_{\sigma(m)}\right)-\cos\left(\theta_{\sigma(n)}\right)
},
\end{align}
where $j_i=2\cos(\theta_i)$ and $S_4$ denotes the set of permutations
of four elements~\footnote{The sum over $S_4$ overcounts equivalent saddle
configurations. For example, the permutations $(1,2,3,4)$ and
$(3,4,1,2)$ give identical contributions. This redundancy produces only
an overall factor, which is absorbed into the normalization.}.

In the edge-scaling regime, the same cubic saddle-point expansion used for
the GOE gives
\begin{align}
Z^U_{\rm edge}
&\propto
e^{-\frac{D}{2}\sum_{i = 1}^4\left(j_i^2 + 2j_i\right)}
\left[
\prod_{m < n}\frac{\partial_{j_m} - \partial_{j_n}}{j_m - j_n}
\right]
\prod_{i = 1}^4\Ai\left(- D^{2/3} j_i\right).
\end{align}
Substitution into the moment formula then yields
Eq.~\eqref{eq:result_edge_GUE}, with algebraic prefactor
$p(x) \approx x^{-5/2} + 7.12\,x^{-7/2} + 11.61\,x^{-9/2} 
 + 8.72\,x^{-11/2} + 3.56\,x^{-13/2} + 0.79\,x^{-15/2} + 0.08\,x^{-17/2}$.

\end{widetext}

\end{document}